\documentclass[conference]{IEEEtran}
\IEEEoverridecommandlockouts
\usepackage{cite}
\usepackage{amsmath,amssymb,amsfonts, amsthm}
\usepackage{algorithm}
\usepackage{algorithmic}
\usepackage{graphicx}
\usepackage{textcomp}
\usepackage{xcolor}
\def\BibTeX{{\rm B\kern-.05em{\sc i\kern-.025em b}\kern-.08em
    T\kern-.1667em\lower.7ex\hbox{E}\kern-.125emX}}
\graphicspath{{./figures/}}

\begin{document}
\title{Leader selection in Vehicular Ad-hoc Networks: a Proactive Approach}

\author{\IEEEauthorblockN{Rusheng Zhang \IEEEauthorrefmark{1}, Baptiste Jacquemot \IEEEauthorrefmark{2}, Kagan Bakirci \IEEEauthorrefmark{2}, Sacha Bartholme \IEEEauthorrefmark{2}, Killian Kaempf \IEEEauthorrefmark{2},  Baptiste Freydt \IEEEauthorrefmark{2}, \\Loic Montandon \IEEEauthorrefmark{2}, Shenqi Zhang \IEEEauthorrefmark{1},  Ozan K. Tonguz \IEEEauthorrefmark{1}}
\IEEEauthorblockA{\IEEEauthorrefmark{1} Department of Electrical and Computer Engineering, Carnegie Mellon University, Pittsburgh, USA \\
\{rushengz, shengqiz\}@andrew.cmu.edu  tonguz@ece.cmu.edu}
\IEEEauthorblockA{\IEEEauthorrefmark{2} School of Computer Science and Communication Systems, Ecole Polytechnique F{\'e}d{\'e}rale de Lausanne, Switzerland
    \\ \{baptiste.jacquemot, kagan.bakirci, sacha.bartholme, killian.kampf, baptiste.freydt, loic.montandon\}@epfl.ch}
}

\maketitle
\begin{abstract}
With the rapid advances in vehicle-to-vehicle (V2V) communications, interest for leveraging V2V communications for different applications has also grown exponentially. Leader selection is an important task required in a number of V2V communications related applications. In this paper, a distributed leader selection algorithm is introduced  for vehicular ad-hoc networks (VANET) over unreliable V2V communications. The algorithm is simple, lightweight, and tightly integrated with  SAE 2735 protocol for vehicular networks. The goal of the algorithm is to provide a robust leader selection method for a group of vehicles to determine a temporary leader in a short time, which could benefit several newly proposed applications based on V2V in traffic control and autonomous driving related areas. 
Simulation results have verified the excellent convergence time, robustness, and other aspects of the algorithm.
\end{abstract}

\begin{IEEEkeywords}vehicular networks, V2V communication, distributed leader selection, connected vehicles, connected automated vehicles, cooperative driving
\end{IEEEkeywords}

\section{Introduction}
In recent years, the rapid advances in vehicle-to-vehicle (V2V) communications has facilitated new applications addressing vehicle safety, traffic efficiency, autonomous driving and other aspects. These applications could potentially improve the driving experience, reduce commute time and fuel consumption, enhance traffic safety, and contribute to other vehicle-related areas in a significant way. For many of these applications, a coordinator vehicle is needed to coordinate the right-of-way among a group of vehicles. Such a coordinator or leader vehicle plays an important role in many applications that aim to achieve a cooperative goal for all vehicles in the group. Viable applications include, but are not limited to, Virtual Traffic Lights (VTL) \cite{ferreira2010self, tonguzred, zhang2018virtual}; intersection management/coordination \cite{elhadef2015adaptable,  lu2018mixed}; on-ramp merging \cite{alston2019automated}; Cooperative Adaptive Cruise Control (CACC) system and platoon maintenance \cite{ jovanovic2005ill}, etc. These applications require vehicles to communicate through unreliable V2V communication channels in a distributed manner and eventually come to a consensus on the selection of leader vehicle.

Therefore, a leader selection scheme over a vehicular ad-hoc network (VANET) is of great interest, especially for connected vehicles (CV) and connected automated vehicles (CAV) related applications. While there exists several results on leader selection in dynamic ad-hoc networks \cite{gupta2000probabilistically, malpani2000leader, vasudevan2003leader, vasudevan2004design, derhab2008self}, these results are not particularly suitable for vehicular networks. One of the crucial aspects is that vehicular network protocols such as SAE 2735\cite{sae20142735} only allow to  periodically broadcast messages to the surrounding vehicles. Such special broadcast features in vehicular networks are  never considered by these algorithms, yet could be particularly helpful for leader selection within a group of small size (typically a size smaller than 40) and should be incorporated into the leader selection process.

In this paper,  a new proactive leader selection algorithm is introduced first. Simulation results are then carried out to show that the algorithm has several desirable features:
\begin{enumerate}
    \item The algorithm is simple and only requires a minimum amount of information sent. Nevertheless, it is very efficient in terms of fast convergence time. The algorithm also achieves fast re-stabilization when uncontrolled erroneous events happen.
    \item The algorithm can be easily integrated to vehicular network protocols \cite{kenney2011dedicated}, it could be implemented by only adding a customized field or header to the Basic Safety Message (BSM).
    \item The algorithm is robust under highly dynamic environments. It does NOT require a static topology and reliable communication channels; it does NOT even require connectivity of the network at all times.
\end{enumerate}
\section{Related Works}
Previously, several leader selection algorithms have been proposed  in a distributed system where processes communicate over unreliable channels. To give some examples,\cite{gupta2000probabilistically} gives a leader selection algorithm for a large group of nodes, assuming random link failure and random node crash, the algorithm is highly scalable, but with only a probabilistic guarantee; \cite{vasudevan2003leader} gives two secure leader selection algorithm for ad-hoc networks, but it assumes static topology during the initialization phase, which could be unrealistic; several leader selection algorithms are proposed for dynamic ad-hoc networks \cite{vasudevan2004design, derhab2008self, malpani2000leader}, these algorithms maintain a spanning tree or a directed acyclic graph (DAG). Hence, they require the nodes to be aware of network connectivity constantly and need to handle node crashes and new network joins explicitly. Therefore, these algorithms are too complicated for local cooperative applications in vehicular networks, considering limited computation power as well as the fact that they are not able to utilize the existing broadcast feature of vehicular networks. Another drawback of the aforementioned algorithms is that these algorithms are strict extrema-finding algorithms that require a leader switch when the order of the nodes changes, which is not a desirable feature for realistic vehicular applications. It is not preferable to switch leaders frequently as it will create gap periods during the process. On the other hand, the leader's order dropping from 'the first' to 'the second' will hardly affect the operation results in most of the applications. Therefore, such leader switches are unnecessary and should be prevented.

Surprisingly, very few leader selection algorithms are designed specifically in the context of vehicular networks applications. \cite{sommer2014networking, fathollahnejad2017probabilistic} proposed a leader-selection algorithm for a specific application in vehicular network, known as Virtual Traffic Lights (VTL), though the algorithm designed is in the context of vehicular networks, the algorithm is still reactive and doesn't take the aforementioned drawbacks into consideration.

The leader selection algorithm introduced in this paper is specifically designed for VANET and bodes well with the vehicular network broadcast feature. The algorithm can be implemented with Dedicated Short Range Communications (DSRC) broadcasting, without any point-to-point communications, which is a highly desirable feature for vehicular network application. This algorithm is also an approximate extrema-finding algorithm, that will select the best leader based on the order function, but will prevent unnecessary leader switches when the order of the nodes changes, which is a desirable feature for most applications, as discussed above.

\section{Problem Statement}
\label{section:problemStatement}
\subsection{Application Requirement}
\label{subsection:applicationRequirement}
We consider a typical VANET scenario for a small group (typically, 40 vehicles maximum) of moving vehicles. All vehicles considered are capable of V2V communications, but the communication channel could be noisy and unreliable, any packet could be lost. Denote $veh_i$ as the i'th vehicle.  An order function $\phi(veh_1, veh_2)$ is given, which gives a strict binary relationship for any two vehicles, it returns 1 or 0 based on the relation of the two vehicles. 
 
 Notice that some papers in the literature prefer height function to order function in extrema-finding tasks. Since this paper is application-oriented, we choose order functions to describe the problem as height functions are not as straightforward as order functions in some vehicular applications. Notice that it is trivial to convert height functions to order functions by directly comparing the height of the two nodes.

With the scenario setting above, the following functionalities are desired:
\begin{enumerate}
    \item The vehicles select a leader based on the order function $\phi$, all the vehicles should come to a consensus and be aware of the same leader. Ideally, the best vehicle should be the leader.
    
    \item The leader status needs to be maintained: When the leader disappears, the new leader selection will start; when new vehicles join the group, they will come to the same consensus. 
    
    \item After selecting the leader, the leader status should be persistent, even if the order of the nodes changes, or a better vehicle joins, as leader switches will create a time gap which is undesirable for most applications.
    
    \item The leader should be able to broadcast uniform information to all vehicles in the group. We denote these message as \textit{leader messages}

\end{enumerate}  
\section{Algorithm}
\label{section:algorithm}
\subsection{The basic proactive leader selection algorithm}
\label{section:basicAlgo}
The leader selection algorithm introduced in this paper is based on proactive broadcasting. All vehicles will broadcast \textit{leader messages} periodically with a period $t_{p}$, to maintain the current leader status. The \textit{leader message} contains information about the current leader, it will only be generated by the leader vehicle itself; all other vehicles will only relay the leader message. The algorithm is simple enough that \textit{leader messages} are the only messages needed to perform leader selection. The leader messages function as information carrier in the leader selection process, as well as a heart-beat indicating that the leader still exists after convergence to one leader. 

\begin{figure}[ht]
    \centering
    \includegraphics[width=.9\linewidth]{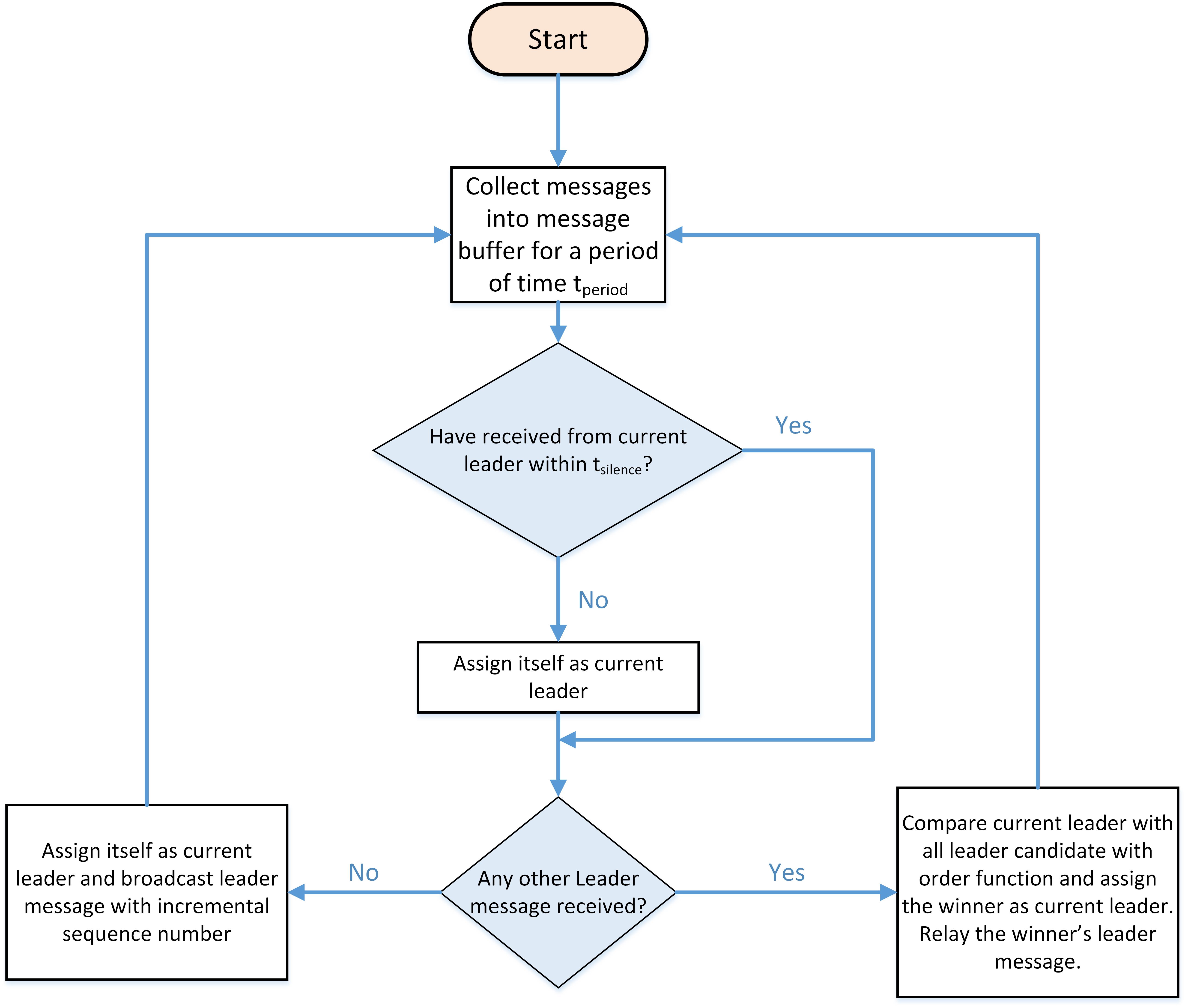}
    \caption{A diagram showing how the basic proactive leader selection algorithm works}
    \label{fig:flowchart}
\end{figure}

The \textit{leader message} is defined to contain the following field:
\begin{itemize}
    \item \textbf{leader:} the leader message specifies the id of the current leader.
    \item \textbf{sequence number:} the unique identifier of each message, so that each message will only be relayed once from each vehicle.
    \item \textbf{information of leader:} This field contains information of the leader in this leader message. This field is used for vehicles to compare leader candidate. It should contain all information needed for order function $\phi$. Most commonly, this field contains GPS information of the vehicle and its unique id to break ties.
\end{itemize}

Figure \ref{fig:flowchart} gives the flowchart of the leader selection process.  Initially, vehicles assign themselves as leader and all of them will issue and broadcast \textit{leader messages} of themselves. Periodically, every vehicle compares the current leader with all the \textit{leader messages} received in the message buffer with the order function $\phi(\cdot)$. If any message contains a better leader than the current leader, the vehicle will replace the current leader with the better leader, and relay that best \textit{leader message} in the buffer. Observe that only the vehicles that consider themselves as leaders issue \textit{leader messages}, other vehicles only relay \textit{leader messages}. In this way, after a certain period of time, all vehicles will arrive at a consensus of one true leader. 

The leader status is maintained by the leader which keeps broadcasting \textit{leader messages} periodically, such \textit{leader messages} are treated as 'heart beat' indicating that the leader is still functional. Each vehicle will check periodically if the heartbeat still exists (by checking if any leader selection message is received within $t_{silence}$). If the heartbeat disappears, the vehicle will reset its leader status by assigning itself as leader and issue \textit{leader messages} again.

Observe that the vehicle will only start to assign the leadership to itself (reset the leader status) when no \textit{leader messages} are received. As long as \textit{leader messages} are received, even if  the vehicle itself becomes the best leader candidate, it will not override leader status. This is a desirable design in leader selection in a vehicular network, as the vehicles are dynamically moving and the GPS signal can be noisy, two vehicles might be racing. This mechanism is designed to stabilize the leader selection process and quickly converge to a stable leader, and to prevent unnecessary frequent leader switches. 


 While the algorithm described above can already fulfill the leader selection task, it can be further optimized to have better bandwidth performance (which is crucial for DSRC communications as the bandwidth is limited). In the remainder of this section, further optimizations of the algorithm are introduced.

\subsection{Preventing broadcast storm}
\label{section:optimizeForBroadcastStorm}
While the aforementioned algorithm already gives good performance, we can further optimize the algorithm to reduce redundant broadcasting in order to save communication bandwidth.

Like all proactive algorithms in ad-hoc networks, the introduced algorithm will generate a large amount of packets and will possibly jam the limited bandwidth in some situations (i.e., the precious bandwidth of DSRC) and cause a broadcast storm \cite{tonguz2006broadcast, wisitpongphan2007broadcast}. This is a known draw-back of all proactive algorithm. 
In this subsection we introduce several mechanism to address this issue.

\subsubsection{Avoid unnecessary relay}
Since the broadcast messages are in the vehicular network scenario, all vehicles will broadcast Basic Safety Message (BSM) periodically, therefore, vehicles will be able to sense their neighbors and maintain a list of all their neighbors, this list will be attached as payload of the leader selection message. Therefore, the vehicle will check the neighbor list of the received leader selection message(s) and the vehicle will only broadcast the leader message if it has a neighbor not in any of the received leader message's neighbor list. This will substantially reduce the broadcast messages within a platoon, especially when most of the vehicles are connected to each other.

\subsubsection{Reduce the broadcast frequency when already reached consensus}
The \textit{leader message} plays different roles before and after reaching consensus. Before reaching a consensus, the \textit{leader messages} are the information carriers, the broadcast frequency of the \textit{leader messages} directly determines the convergence time. High-frequency leader messages broadcast will make vehicles come to consensus quickly, hence it is important to broadcast leader messages frequently.  After reaching consensus, the leader messages are used as heartbeat to indicate that the leader is still functional, it is not critical to keep a high broadcast frequency in this case. Therefore, for the leader vehicle, as long as it doesn't receive the conflict leader messages, it will broadcast leader message at a lower rate. This mechanism will make high-frequency broadcasting only happen during the leader selection process, or whenever a new vehicle joins. 

It is worth mentioning that for some applications, leader messages are also used for the leader to broadcast application information to other vehicles, in the case that application information requires a high refreshing frequency, the leader shouldn't reduce the broadcast frequency.

\section{Simulation Evaluation}
\subsection{Simulation scenario}
We performed extensive packet-level simulations to evaluate the performance of the designed algorithm in different settings. To generate reliable and realistic results, we developed a hybrid simulator that simulates both the mobility of vehicles and probabilistic DSRC channel. The mobility is simulated using SUMO, a popular open-source mobility simulator \cite{SUMO2018}. 

As for the communication channel modeling, to yield realistic results, proper probability model for packet reception rate (PRR) is adopted. Previous studies have proposed the Nakagami-m distribution for DSRC channel modeling \cite{killat2009empirical}, the PRR can be obtained by the following equation:

\begin{equation} 
    P_R(d,CR) = e^{-m(d/CR)}\sum_{i=1}^{m}\frac{(m(d/CR))^{i-1}}{(i-1)!}
    \label{eq:proba_packet}
\end{equation}

In the equation, $m$ is the fading parameter of the signal. It has different values due to the weather, the congestion of the network or the number of buildings. This parameter varies from 1 to 3, $m = 1$ corresponds to a harsh communication condition, and $m=3$ corresponds to an ideal communication condition. 
 Parameter $d$ corresponds to the distance between the two vehicles. $CR$ corresponds to the intended communication range by the radio \cite{azimi2013reliable}. This parameter is determined by the radio transmission power. As 802.11p specifies five power levels, $CR$ can take five different values: 100, 200, 300, 400, 500  \cite{killat2009empirical} which coorespond to the 5 power levels of DSRC: 5, 10, 15, 20 and 33 dBm. For DSRC OnBoard Units (OBU), $CR=100$ is normally used.




We performed the simulations for a typical one-lane intersection. The length of each approach (i.e., block length)is 100 meters. A traffic light is placed at the intersection that performs periodical phase changes. Vehicles approaching the intersection will run the leader selection algorithm described above. When the leader vehicle passes the intersection, the leader will disappear and all other vehicles need to detect this event and select a new leader immediately. Vehicles arrive at the intersection according to a Poisson process, for convenience, the arrival rate of 4 approaches are set to be the same.
 
Two algorithms are evaluated, the \textbf{basic leader selection algorithm} introduced in section \ref{section:basicAlgo} as well as the \textbf{optimized leader selection algorithm} that applies all optimization methods introduced in \ref{section:optimizeForBroadcastStorm}. The simulation code and instructions on how to use the code can be found in \cite{leaderselectionrepo}.

\subsection{Qualitative results}
\label{section:qualitative}
We first visualize the leader status in each SUMO simulation in order to obtain a qualitative understanding of the algorithm.  The leader will stop at the intersection for a while waiting for the red light and leave the intersection when the light turns green. We observe that at the moment the leader vehicle leaves the intersection, all other vehicles at the intersection detect this event automatically and start selecting a new leader, this  new leader selection period is very short (less than 1 second). In terms of leader status maintenance, the performance of the \textbf{basic leader selection algorithm} is almost identical to the performance of the \textbf{optimized leader selection algorithm}  (but there is a distinct difference between the two algorithms in terms of message exchange, more details can be found in section \ref{section:quantitative}).

\begin{figure}[ht]
    \centering
    \includegraphics[width=.9\linewidth]{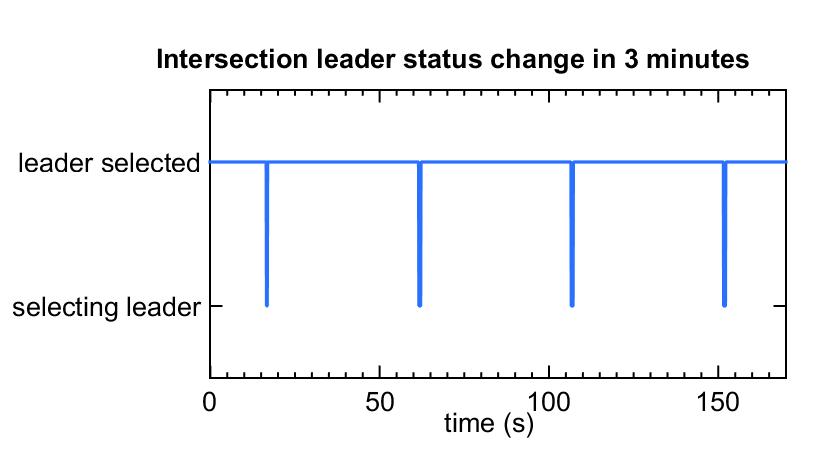}
    \caption{A typical simulation of 3 minutes using the leader selection algorithm described in this paper.}
    \label{fig:leaderStatus}
\end{figure}

Figure \ref{fig:leaderStatus} shows the leader status timeline of one typical simulation. The status 'leader selected' is the status where the algorithm successfully selected one leader at the intersection, the status 'selecting leader' is the status where the algorithm has not converged yet and is still in the process of selecting a leader. Because that leader vehicle will periodically leave the intersection, the leader selection algorithm will need to re-select a leader periodically, the re-selecting process can be clearly observed and located by the spikes in the figure. These spikes are the very short period of the leader selection process. In the figure \ref{fig:leaderStatus}, the width of the spikes can't be observed clearly because they are very short, quantitative measurements are given in \ref{section:quantitative}. Figure \ref{fig:leaderStatus} and the simulation observation results qualitatively show that the leader selection algorithm works as expected in a realistic vehicular application scenario.

\subsection{Quantitative results}
\label{section:quantitative}

In this subsection, we present quantitative measurements which show the performance of the algorithm in different typical scenarios. The following metrics are chosen to illustrate the performance. These statistical metrics are collected from 100 simulations, each of which lasts 3 minutes:
\begin{itemize}
\item \textbf{Stable percentage}: Percentage of time when there is a unique leader at the intersection, it's the \textit{average} value over all simulations.
\item \textbf{Average convergence time}: Time needed to converge to a unique leader, it's the \textit{average} taken over all simulation trials. 
\item \textbf{Maximum convergence time}: Time needed to converge to a unique leader, it's the \textit{average} of all simulation trials. 
\item \textbf{Number of messages} sent in each simulation. We use the  \textit{average} number of messages sent over all simulation trials. This will give the performance of the algorithm in the channel usage aspect.
\end{itemize} 
\begin{table}[ht]
\caption{Performance evaluation of the leader selection algorithm}
\begin{tabular}{|l|c|c|c|c|}
\hline
Algorithm & \multicolumn{2}{l|}{Basic} & \multicolumn{2}{l|}{Optimized} \\ \hline
traffic volume & \multicolumn{1}{l|}{medium} & \multicolumn{1}{l|}{dense} & \multicolumn{1}{l|}{medium} & \multicolumn{1}{l|}{dense} \\ \hline
Stable percentage & 97\% & 98\% & 97\% & 98\% \\ \hline
Average convergence time (s) & 0.66 & 0.6 & 0.51 & 0.39 \\ \hline
Maximum convergence time (s) & 0.88 & 0.83 & 0.91 & 0.64 \\ \hline
Number of messages & 13474 & 54743 & 5080 & 8829 \\ \hline
\end{tabular}
\label{tab:performance}
\end{table}

Table \ref{tab:performance} shows the results obtained from  the simulation. From the table, we observe several interesting facts. The stable percentage, i.e., the percentage of the time that all vehicles are having the same leader, is in general very high (97\% - 98\%). the remaining 2-3\% are the duration of leader selection process. This can be justified from the leader status figure in Figure \ref{fig:leaderStatus}.

As for the convergence time, i.e., the average convergence time is roughly half a second, and the maximum convergence time is less than 1 second in all scenarios. This is an ideal performance as it is critical for most vehicular applications that the leader selection process converges fast. An average of half a second and maximum less than a second is sufficient for most of vehicular applications.

Another interesting observation is that the traffic volume does not affect stable percentage and convergence time in a major way. This seems counter-intuitive at a first glimpse, as when vehicles number increases, more vehicles need to come to an agreement. The reason is that when there are more vehicles, the \textit{leader message} will be broadcast more times by each vehicle, hence increasing the probability for each vehicle to receive the \textit{leader message}. 

Meanwhile, we notice that, in comparison to the basic algorithm, the optimized algorithm will reduce the amount of messages in a significant way. It reduces 60\% of messages in medium traffic and 85\% in dense traffic. It is desirable that optimized algorithm reduces a larger percentage of the messages in dense traffic, because the goal of the optimized algorithm is to alleviate the broadcast storm, especially under dense traffic conditions.
\section{Discussion}

One of the future works is to carry out a  more detailed mathematical analysis of the performance as well as a strict proof of correctness of the algorithm. A more comprehensive scheme that specifies leader switches as well as the messages from all member nodes to the leader will also be investigated in the future.

\section{Conclusion}
In this paper, a new leader selection algorithm is introduced for vehicular ad-hoc networks. The algorithm is based on proactive broadcasting, which can be easily integrated into the SAE 2735 protocol for vehicular applications. 

Simulation results have shown that the algorithm converges to a unique leader in very short time (within 1 second). When using the algorithm at an intersection,  vehicles reach consensus on one leader during 98\% of the time. The remaining 2\% of the time is the time spent on leader switching; all leader switchings take less than 1 second.




%
\bibliographystyle{IEEEtran}
\bibliography{ref}  
\end{document}